\documentclass[12pt]{iopart}
\usepackage{iopams}

\newcommand{\NO}{\nonumber\\}
\newcommand{\Ave}[1]{\left< #1 \right>_{\tau}}
\newcommand{\Avet}[1]{\left< #1 \right>_{t}}

\newcommand{\Poixs}[2]{\left\{ #1 ~,~ #2 \right\}_{\bxs}}
\newcommand{\Poix}[2]{\left\{ #1 ~,~ #2 \right\}_{\bx}}
\newcommand{\Poic}[2]{\left\{ #1 ~,~ #2 \right\}_{\bchi}}
\newcommand{\ep}{\epsilon}
\newcommand{\Od}[1]{O(\epsilon^{#1})}
\newcommand{\Hs}[1]{H^{\ast}_{#1}}
\newcommand{\HRG}{H^{\scriptsize \mbox{RG}}}
\newcommand{\fracd}[2]{\frac{\rmd #1}{\rmd #2}}
\newcommand{\fracp}[2]{\frac{\partial #1}{\partial #2}}
\newcommand{\fracpp}[3]{\frac{\partial^{2} #1}{\partial #2 \partial #3}}
\newcommand{\sumres}[1]{\sum_{#1 :\mbox{\scriptsize Res}}}
\newcommand{\sumnon}[1]{\sum_{#1 :\mbox{\scriptsize Non}}}
\newcommand{\Hh}[1]{\hat{H}_{1,#1}}

\newcommand{\bI}{\boldsymbol{I}}
\newcommand{\bx}{\boldsymbol{x}}
\newcommand{\bxs}{\boldsymbol{x}^{\ast}}
\newcommand{\bIs}{\boldsymbol{I}^{\ast}}
\newcommand{\bTs}{\btheta^{\ast}}
\newcommand{\bX}[2]{\boldsymbol{x}_{#1}^{(#2)}}
\newcommand{\bIe}[1]{\boldsymbol{I}^{(#1)}}
\newcommand{\bTe}[1]{\btheta^{(#1)}}
\newcommand{\bm}{\boldsymbol{m}}
\newcommand{\bn}{\boldsymbol{n}}
\newcommand{\momega}{\bm\cdot\bomega}
\newcommand{\nomega}{\bn\cdot\bomega}
\newcommand{\bA}{\boldsymbol{A}}
\newcommand{\bB}{\boldsymbol{B}}
\newcommand{\bC}{\boldsymbol{C}}

\begin{document}

\begin{flushright}
  DPNU-99-05
\end{flushright}

\jl{1}

\letter{Renormalization group method and canonical perturbation theory}

\author{Yoshiyuki Y. YAMAGUCHI\S
\footnote[1]{e-mail: yamaguch@se.ritsumei.ac.jp}
  and Yasusada NAMBU\P
\footnote[2]{e-mail: nambu@allegro.phys.nagoya-u.ac.jp} }

\address{\S\ The general research organization of science and
  engineering, Ritsumeikan University,  
  Kusatsu, 525-8577, Japan}
\address{\P\  Department of Physics, 
  Nagoya  University, Nagoya 464-8602, Japan}


\begin{abstract}
  Renormalization group method is one of the most powerful tool 
  to obtain approximate solutions to differential equations.
  We apply the renormalization group method to Hamiltonian systems
  whose integrable parts linearly depend on action variables.
  We show that the renormalization group method gives 
  the same approximate solutions as canonical perturbation theory 
  up to the second order of a small parameter 
  with action-angle coordinates.
\end{abstract}


\submitted


\section{Introduction}
\label{sec:introduction}

Dynamical systems written by differential equations
are useful to understand temporal evolutions of the nature.
Exact solutions to the equations are not always obtained
because of non-integrability of systems,
and naive perturbation often yields secular terms
which prevent us from getting approximate but global solutions.
Singular perturbation techniques~\cite{hinch-91},
eg averaging methods, multi scale methods, 
matched asymptotic expansions and WKB methods,
are available to construct global solutions.
However, they provide no systematic procedures for general systems
because we must select a suitable assumption
about the structure of a perturbation series.

Recently, renormalization group method is proposed 
~\cite{chen-96,kunihiro-95}
as a tool for global asymptotic analysis of the solutions to 
differential equations.
It unifies the techniques listed above, 
and can treat many systems irrespective of their features.
We apply the renormalization group method to Hamiltonian systems,
and compare it with canonical perturbation theory~\cite{hori-66,deprit-69},
which is one of the most developed perturbation theory 
for Hamiltonian systems.
In this letter, we show that the renormalization group method 
also unifies the canonical perturbation theory.
That is, the former and the latter give the same solutions
to equations of motion up to the second order of a small parameter.

We use action-angle coordinates
as they are suitable for perturbed Hamiltonian systems,
and Hamiltonians are
\begin{equation}
  \label{eq:model}
  H(\bI,\btheta) = H_{0}(\bI) + \ep H_{1}(\bI,\btheta),
\end{equation}
where both $\bI$ and $\btheta$ are $N$-dimensional vectors,
the integrable part $H_{0}$ is
\begin{equation}
  \label{eq:0th-Hamiltonian}
  H_{0}(\bI) = \bomega \cdot \bI ,
\end{equation}
and $H_{1}(\bI,\btheta)$ is periodic 
with respect to each element of $\btheta$.

We derive an approximate solution with naive perturbation 
in section \ref{sec:naive-sol},
and then we renormalize secular terms to constants of integration
in section \ref{sec:renormalization}.
Finally, in section \ref{sec:CP}
we compare the renormalized solutions with
solutions obtained by canonical perturbation theory.

\section{Naive Solution}
\label{sec:naive-sol}

The equation of motion for the system (\ref{eq:model}) is
\begin{equation}
  \label{eq:eom}
  \fracd{\bx}{t} = \Poix{\bx}{H_{0}(\bx) + \ep H_{1}(\bx)} , 
\end{equation}
where $\bx=(\bI,\btheta)$ is a $2N$-dimensional vector 
and the symbol $\Poix{\cdot}{\cdot}$ is Poisson bracket
with respect to the subscript, in this case, $\bx$.
We expand $\bx$ as a series of powers of $\ep$,
\begin{equation}
  \label{eq:power-series}
  \bx = \bX{}{0} + \ep \bX{}{1} + \ep^{2} \bX{}{2} + \dots , 
\end{equation}
and then equation (\ref{eq:eom}) becomes
\begin{equation}
  \hspace*{-4em}
  \fracd{}{t} \left( 
    \bx^{(0)} + \ep \bx^{(1)} + \ep^{2} \bx^{(2)} + \dots
  \right)
  = \Poix{\bx}{
    H_{0} + \ep H_{1}
    + \ep^{2} \fracp{H_{1}}{\bx} \cdot \bX{}{1}
    } (\bX{}{0})
  + \dots ,
\end{equation}
where, in the right-hand-side, we substitute $\bX{}{0}$ to $\bx$
after the Poisson bracket has been operated.

The solution to $\Od{0}$ is
\begin{equation}
  \bIe{0} = \balpha_{0} , \qquad
  \bTe{0} = \bomega\ t + \bbeta_{0} ,
\end{equation}
where $N$-dimensional vectors $\balpha_{0}$ and $\bbeta_{0}$ are
constants of integration.

The equation of motion for $\Od{1}$ is
\begin{equation}
  \fracd{\bx^{(1)}}{t} = \Poix{\bx}{H_{1}}(\bX{}{0}) ,
\end{equation}
and hence the solution to $\Od{1}$ is
\begin{equation}
  \eqalign{
    \bX{}{1} 
    & = \Poic{\bchi}{S_{1}(\bchi)} 
    +\ t~\Poic{\bchi}{\Avet{H_{1}(\bchi)}} ,
    } 
\end{equation}
where we introduced the symbols
\begin{equation}
  \label{eq:chi}
    \bchi = (\balpha,\bbeta) , \qquad
    \balpha = \balpha_{0}, \qquad
    \bbeta  = \bomega\ t + \bbeta_{0} ,
\end{equation}
$\Avet{\cdot}$ represents average over $t$, and
\begin{equation}
  \label{eq:S_1}
  S_{1}(\bchi) \equiv \int \rmd t\ ( H_{1}(\bchi) - \Avet{H_{1}(\bchi)} ) .
\end{equation}
The following relation was also used 
\begin{equation}
  \Poix{f(\bx)}{g(\bx)}\ (\bX{}{0}) 
  = \Poic{f(\bchi)}{g(\bchi)} ,
\end{equation}
which is satisfied by arbitrary functions $f$ and $g$
that are periodic for $\btheta$ and $\bbeta$.

The equation to $\Od{2}$ is 
\begin{equation}
  \eqalign{
    \fracd{\bx^{(2)}}{t}
    & = \Poic{\bchi}{ \Poic{H_{1}(\bchi)}{S_{1}(\bchi)} }
    + \Poic{ \Poic{\bchi}{S_{1}(\bchi)} }{H_{1}(\bchi)} \\
    & \hspace*{3em} 
    + t~\Poic{ \Poic{\bchi}{H_{1}(\bchi)} }{\Avet{H_{1}(\bchi)}} ,
    } 
\end{equation}
and the solution to $\Od{2}$ is
\begin{equation}
  \label{eq:x_2}
  \eqalign{
    \bX{}{2}
    & = \Poic{\bchi}{S_{2}} 
    + \frac{1}{2} \Poic{ \Poic{\bchi}{S_{1}} }{S_{1}}
    + t~\Poic{\bchi}{\Avet{F_{2}}} \\
    & \hspace*{2em}
    + (t^{2} \mbox{-secular terms})
    + (t \mbox{-secular terms with non-constants}) .
  }
\end{equation}
Here
\begin{equation}
  \eqalign{
    \label{eq:S_2}
    S_{2}(\bchi) & \equiv \int \rmd t\ 
    ( F_{2}(\bchi) - \Avet{F_{2}(\bchi)} ), \\
    \label{eq:F_2}
    F_{2}(\bchi) & \equiv \Poic{H_{1}(\bchi)}{S_{1}(\bchi)} 
    + \frac{1}{2} \Poic{ \Poic{H_{0}(\bchi)}{S_{1}(\bchi)} }{S_{1}(\bchi)} ,
    }
\end{equation}
and 
\begin{eqnarray}
  \hspace*{-3em} & \Poic{ \Poic{\bchi}{S_{1}} }{H_{1}}
  = \Poic{\bchi}{ \frac{1}{2} \Poic{ \Poic{H_{0}}{S_{1}} }{S_{1}} }
  + \bA(\bchi) + \bB(\bchi) , \\
  \hspace*{-3em} & \int \rmd t\ \bA(\bchi) = \frac{1}{2} 
  \Poic{ \Poic{\bchi}{S_{1}} }{S_{1}} , \\
  \hspace*{-3em} & \int \rmd t\ t~\Poic{ \Poic{\bchi}{H_{1}} }{\Avet{H_{1}}} 
  = - \int \rmd t\ \bB(\bchi) \NO
  \hspace*{-3em} & \hspace*{3em}
  + (t^{2} \mbox{-secular terms})
  + (t \mbox{-secular terms with non-constants}) ,
\end{eqnarray}
which are proven by using Fourier expressions
of $H_{1}(\bchi), \Avet{H_{1}(\bchi)}$ and $S_{1}(\bchi)$.
The concrete forms of the Fourier expressions,
$\bA(\bchi)$ and $\bB(\bchi)$ are shown in appendix.

Consequently, the naive solution to equation (\ref{eq:eom}) is,
up to $\Od{2}$,
\begin{equation}
  \label{eq:naive-sol}
  \eqalign{
    \bx & = \bchi + \ep \Poic{\bchi}{S_{1}}
    + \ep^{2}
    \left[
      \Poic{\bchi}{S_{2}}
      + \frac{1}{2} \Poic{ \Poic{\bchi}{S_{1}} }{S_{1}}
    \right] \\
    & \hspace*{1em} + t~
    \left[
      \ep \Poic{\bchi}{\Avet{H_{1}}}
      + \ep^{2} \Poic{\bchi}{\Avet{F_{2}}}  
    \right] \\
    & \hspace*{1em} + \ep^{2} \left[
    (t^{2} \mbox{-secular terms})
    + (t \mbox{-secular terms with non-constants}) 
    \right] .
    }
\end{equation}

\section{Renormalization of Secular Terms}
\label{sec:renormalization}

We renormalize the secular terms of the naive solution (\ref{eq:naive-sol})
to the constants of integration.
First we regard $\balpha_{0}$ and $\bbeta_{0}$ as functions of $t$ which are
\begin{equation}
  \label{eq:RG-trans}
  \eqalign{
    \balpha_{0}(t) & = \balpha_{0} + t \left[
      \ep \Poic{\balpha}{\Avet{H_{1}}}
      + \ep^{2} \Poic{\balpha}{\Avet{F_{2}}}  
    \right] , \\
    \bbeta_{0}(t) & = \bbeta_{0} + t \left[
      \ep \Poic{\bbeta}{\Avet{H_{1}}}
      + \ep^{2} \Poic{\bbeta}{\Avet{F_{2}}}  
    \right] .
    }
\end{equation}
Second we introduce assumptions with which 
the renormalization transformations (\ref{eq:RG-trans})
becomes to be a Lie group.
In this case, we assume that equation (\ref{eq:RG-trans})
is a truncated Taylor series of $\balpha_{0}(t)$ and $\bbeta_{0}(t)$
around the initial time $t=0$ ~\cite{matsuba-98} .
From time-evolutional symmetry of the system (\ref{eq:model}),
the renormalization group equation becomes
\begin{equation}
  \label{eq:RGE-albe}
  \eqalign{
    \fracd{\balpha_{0}}{t}
    & = \ep \Poic{\balpha}{\Avet{H_{1}}}
    + \ep^{2} \Poic{\balpha}{\Avet{F_{2}}} + \Od{3} , \\
    \fracd{\bbeta_{0}}{t}
    & = \ep \Poic{\bbeta}{\Avet{H_{1}}}
    + \ep^{2} \Poic{\bbeta}{\Avet{F_{2}}} + \Od{3} ,
    } 
\end{equation}
in other words,
\begin{equation}
  \label{eq:RGE}
  \fracd{\bchi}{t} = \Poic{\bchi}{H_{0}(\bchi)}
  + \ep \Poic{\bchi}{\Avet{H_{1}}}
  + \ep^{2} \Poic{\bchi}{\Avet{F_{2}}} + \Od{3} .
\end{equation}
The renormalized solution is therefore
\begin{equation}
  \label{eq:renormalized-sol}
  \hspace*{-3em}
  \bx = \bchi + \ep \Poic{\bchi}{S_{1}}
  + \ep^{2}
  \left[
    \Poic{\bchi}{S_{2}}
    + \frac{1}{2} \Poic{ \Poic{\bchi}{S_{1}} }{S_{1}}
  \right] + \Od{3} ,
\end{equation}
where $\bchi$ is governed by equation (\ref{eq:RGE}).
Here, $t^{2}$-secular terms and 
$t$-secular terms with non-constants in $\Od{2}$
of equation (\ref{eq:naive-sol})
are renormalized to coefficients of $t$-secular terms
and coefficients of non-constant terms, respectively.

\section{Comparison with Canonical Perturbation Theory}
\label{sec:CP}

Finally we compare the renormalized solution (\ref{eq:renormalized-sol})
and solution obtained by canonical perturbation theory.
The strategy of the theory is to canonically transform
coordinates $\bx=(\bI,\btheta)$ to $\bxs=(\bIs,\bTs)$ 
with the generator $S(\bxs)$ 
\begin{equation}
  \label{eq:bxtobxs}
  \bx = \exp(\ep D_{S}) \bxs, \qquad
  D_{S}f(\bxs) \equiv \Poixs{f(\bxs)}{S(\bxs)} ,
\end{equation}
such that secular terms do not appear in the coordinates $\bxs$.
What we must calculate are the generator $S$ 
and the transformed Hamiltonian $\Hs{}$ .
Canonical perturbation theory ~\cite{hori-66,deprit-69} states that 
the required generator $S=S_{1}+\ep S_{2}+\dots$ is expressed as
\begin{eqnarray}
  \label{eq:generator-S1}
  S_{1}(\bxs) & = \int \rmd\tau\ ( H_{1}(\bxs) -\Ave{H_{1}(\bxs)} ) , \\
  \label{eq:generator-S2}
  S_{2}(\bxs) & = \int \rmd\tau\ ( F_{2}(\bxs) -\Ave{F_{2}(\bxs)} ) ,
\end{eqnarray}
and the transformed Hamiltonian 
$\Hs{}=\Hs{0}+\ep\Hs{1}+\ep^{2}\Hs{2}+\dots$ as
\begin{eqnarray}
  \label{eq:Hs}
  \Hs{0}(\bxs) & = H_{0}(\bxs) , \quad
  \Hs{1}(\bxs) = \Ave{ H_{1}(\bxs) } , \quad
  \Hs{2}(\bxs) = \Ave{ F_{2}(\bxs) } , 
\end{eqnarray}
where
\begin{equation}
  \label{eq:F_2_CP}
  \hspace*{-2em}
  F_{2}(\bxs) = \Poixs{H_{1}(\bxs)}{S_{1}(\bxs)} 
  + \frac{1}{2} \Poixs{ \Poixs{H_{0}(\bxs)}{S_{1}(\bxs)} }{S_{1}(\bxs)} .
\end{equation}
The symbol $\Ave{\cdot}$ represents the average over $\tau$,
the time of $\bxs$ following $\Hs{0}$, that is,
\begin{equation}
  \fracd{\bxs}{\tau} = \Poixs{\bxs}{\Hs{0}(\bxs)} .
\end{equation}

Consequently, this theory gives an approximate solution 
determined by
\begin{equation}
  \label{eq:moe-xs}
  \hspace*{-1em}
  \fracd{\bxs}{t} 
  = \Poixs{\bxs}{H_{0}(\bxs)}
  + \ep \Poixs{\bxs}{\Ave{H_{1}}}
  + \ep^{2} \Poixs{\bxs}{\Ave{F_{2}}} 
  + \Od{3}
\end{equation}
and the canonical transformation (\ref{eq:bxtobxs}) 
\begin{equation}
  \label{eq:cano-trans}
  \hspace*{-3em}
  \bx = \bxs + \ep \Poixs{\bxs}{S_{1}} 
  + \ep^{2} \left[
    \Poixs{\bxs}{S_{2}} + \frac{1}{2} \Poixs{\Poixs{\bxs}{S_{1}}}{S_{1}}
  \right] + \Od{3} .
\end{equation}
The approximate solution (\ref{eq:cano-trans}) 
is the same as the renormalized solution (\ref{eq:renormalized-sol})
since temporal evolutions of $\bxs$ and $\bchi$
are governed by equations (\ref{eq:moe-xs}) and (\ref{eq:RGE})
respectively, and the two equations have the same structure.

\section{Summary and Discussions}
\label{sec:summary}

We showed that renormalization group method
gives the same approximate solutions as canonical perturbation
theory up to the second order of a small parameter 
to the Hamiltonian systems 
whose integrable parts linearly depend on action variables.
That is, renormalization group method unifies
not only averaging methods, multi scale methods, 
matched asymptotic expansions and WKB methods,
but canonical perturbation theory.
We suppose that the unification holds even in higher orders 
of the small parameter.

In systems whose integrable parts are not linear,
secular terms are not always proportional to time $t$,
and may be proportional to $t^{n}\ (n\neq 1)$.
Canonical perturbation theory cannot remove the latter secular terms
since subtracting time-averages of perturbative part of Hamiltonian
is effective only for the $t$-linear secular terms.
On the contrary, renormalization group method gives global solutions
by introducing assumptions with which 
renormalization transformations become to a Lie group ~\cite{nambu-99}
and can treat $t^{n}$ type secular terms.

In the previous paper ~\cite{yamaguchi-98},
we discussed relation between integrability of original systems
and symplectic properties of renormalization group equations
in Cartesian coordinates.
From equation (\ref{eq:RGE}),
we clarified that renormalization group equations
are always Hamiltonian systems in action-angle coordinates
whose Hamiltonian is
\begin{equation}
  \label{eq:HRG}
  \HRG(\bchi) = H_{0} + \ep \Avet{H_{1}} + \ep^{2} \Avet{F_{2}}.
\end{equation}
Symplectic properties are recovered 
even in Cartesian coordinates by using ``gauge freedom'' 
which is homogeneous terms of $\Od{1}$.
Details will show in the next paper~\cite{nambu-99}.

YN is supported in part by the Grand-In-Aid 
for Scientific Research of the Ministry of Education, Science, Sports and 
Culture of Japan(09740196). 



\appendix
\section*{Appendix}
\label{sec:appendix}

Let us introduce Fourier series of 
$H_{1},\Avet{H_{1}}$ and $S_{1}$ as
\begin{eqnarray*}
  H_{1}(\bchi) & = \sum_{m} \Hh{\bm}(\balpha) e^{\rmi\bm\cdot\bbeta} , \\
  \Avet{H_{1}}(\bchi) & = \sumres{\bm} \Hh{\bm}(\alpha) 
  \rme^{\rmi\bm\cdot\bbeta} . \\ 
  S_{1}(\bchi) & = \sumnon{\bm} \frac{1}{i\momega} 
  \Hh{\bm}(\balpha) \rme^{\rmi\bm\cdot\bbeta} .
\end{eqnarray*}
where the symbols $\sumres{\bm}$ and $\sumnon{\bm}$ represent
to take summations over $\bm$ 
such that $\momega=0$ and $\momega\neq 0$, respectively.
By using these expressions and $\bfeta=(-\btheta,\bI)$,
the concrete forms of $\bA(\bchi)$ and $\bB(\bchi)$ are
\[
   \bA(\bchi)
   = \frac{1}{2} \sumnon{\bm} \sumnon{\bn} 
   \left(
     \frac{1}{\rmi\momega} + \frac{1}{\rmi\nomega}
   \right) 
   \bC(\bchi)
\]
and 
\[
    \bB(\bchi)
    = \sumres{\bm} \sumnon{\bn}
    \frac{1}{\rmi\nomega} 
    \bC(\bchi)
\]
respectively, where
\[
  \eqalign{
    \bC(\bchi) 
    & =
    \left[
      \rmi n_{k}
      \left(
        \fracp{\Hh{\bn}}{\bfeta} 
        + \fracp{(\rmi\bn\cdot\bbeta)}{\bfeta} \Hh{\bn}
      \right)
      \fracp{\Hh{\bm}}{\alpha_{k}}
    \right. \\
    & \hspace*{2em} 
    - \left.
      \rmi m_{k}
      \left(
        \fracpp{\Hh{\bn}}{\alpha_{k}}{\bfeta} 
        + \fracp{(\rmi\bn\cdot\bbeta)}{\bfeta} \fracp{\Hh{\bn}}{\alpha_{k}}
      \right)
      \Hh{\bm}
    \right]
    \rme^{\rmi (\bm+\bn)\cdot\bbeta} .
    }
\]

\end{document}